\documentclass[preprint]{aastex}

\slugcomment{To appear in Astrophys J. March 10, 2002}

\shorttitle{Djorgovski et al.}
\shortauthors{Collapsed Cores in Globular Clusters}

\begin{document}


\title{Ambipolar Drift in a Turbulent Medium}


\author{Ellen G. Zweibel}
\affil{JILA \& Department of Astrophysical \& Planetary Sciences,
University of Colorado, Boulder CO 80309}



\begin{abstract}
The interstellar magnetic field strength and density are observed to be
correlated, but there is a large
dispersion in this
relation. In particular, the magnetic field
is often observed to be weaker than expected. At
low ionization fraction,
ion-neutral drift, or ambipolar
diffusion, permits slip of the field relative to the neutral gas
and tends to make the field strength more uniform, but is thought to be too
slow to explain the observations.
The purpose of this paper is to show that ion-neutral drift is significantly
faster in a turbulent medium than in a quiescent one. We suggest that this fast ambipolar diffusion can explain
the surprisingly low magnetic fieldstrengths sometimes observed in dense
interstellar gas.
\end{abstract}




\section{Introduction}

The interstellar magnetic fieldstrength and gas density are observed to be
correlated (Troland \& Heiles 1986, Crutcher 1999). This
relationship is thought to arise from
so-called ideal magnetohydrodynamic (MHD) processes, in
which the field
is frozen to the ambient medium.
Under ideal MHD conditions, if the ratio of
mass to magnetic flux were everywhere constant, the slope $q\equiv
d\log B/d\log\rho$ of the
fieldstrength - density correlation would be unity for compression normal to
$\mathbf{B}$, zero for compression parallel to $\mathbf{B}$, and 2/3 for
isotropic compression. The observed value of $q$ is approximately 0.5, which is
consistent with equilibrium models of self gravitating clouds which evolved
under conditions of frozen flux (Mouschovias 1976, Tomisaka, Ikeuchi, \&
Nakamura 1988).

However, a number of observations and upper limits on magnetic
fieldstrength suggest that $B\propto\rho^{0.5}$ is more an upper
envelope than a scaling law. This is true both in atomic and molecular gas
(Bourke et al. 2001, Crutcher 1999, Heiles 2001a, Heiles \& Troland 2001)
Although the number of measurements is small, and the field may be
underresolved in some cases (Brogan \& Troland 2001), the trend towards
weak fields is clear. A different line of
argument comes from numerical simulations of molecular clouds: Padoan \&
Nordlund (1999) claim that models with weak fields replicate the observations
better than models with strong fields.

Weak fields are difficult to reconcile with ideal MHD.
This is particularly so in the case of turbulent molecular
clouds. Giant molecular clouds (GMCs) are about 40 times more dense than the
mean
interstellar gas, suggesting that the magnetic field should be 6-7 times
stronger than the mean field. If GMCs are formed by flows parallel to
the field then of course the field would not be strengthened, but the
mean field is not strong enough to
resist compression and collimate the flows.
Furthermore, Mestel (1985) has pointed out that
accumulation of the mass of a GMC by compression in one dimension requires
organized motion of the gas over nearly one kiloparsec, the origin of which is
difficult to understand.

Thus, we
seek an explanation for the weakness of the
magnetic fieldstrength - gas density correlation beyond the scope
of ideal MHD. At the very largest lengthscales,
ideal MHD should be an excellent approximation.
Moving downwards in scale, the first non-ideal effect
encountered, at least at low ionization fraction,
is ion-neutral drift, or ambipolar diffusion. At the ambipolar
scale, which is many orders of magnitude larger than the resistive scale, the
magnetic field and plasma become decoupled from the neutral material. This
makes it possible to change the mass to flux ratio without altering the
magnetic topology. Ambipolar drift has been invoked as the primary magnetic
flux transport mechanism in dense, star forming gas since the classic paper
by Mestel \& Spitzer (1956). However, it is thought to be too slow to be an
effective transport mechanism in diffuse gas (see \S 2 for quantitative
estimates).

Interstellar gas is turbulent. Turbulent diffusion of
quantities such as heat and angular momentum is often invoked in astrophysics
as a mechanism for enhancing transport rates above
their kinetic theory values, which are usually very slow. Turbulence
enhances diffusion rates by mixing the
relevant quantity to the small scales at which
molecular diffusion operates. This
leads to a mixing time which is approximately the eddy turnover time, and is
nearly independent of the molecular diffusivity.

Whether turbulence enhances the resistive decay rate of a magnetic field is
unclear,
because there is substantial evidence that magnetic
forces resist stretching the field sufficiently to mix it to the tiny scales
at which resistivity operates (Cattaneo
\& Vainshtein 1991, Cattaneo 1994, Cattaneo, Hughes, \& Kim 1996). This paper
addresses a different question, namely, whether turbulence in a weakly ionized
gas can transport magnetic field with respect to the neutral matter, without
resistive dissipation necessarily coming into play.
Because the ambipolar drift scale is much larger than the resistive scale, the
feedback effects which can suppress turbulent resistivity are far less dramatic,
although they cannot always be ignored.

We use analytical methods to calculate the mixing rate. Numerical
study of mixing
requires that numerical diffusion of both field and fluid  be very
well controlled. Analytical calculations are useful for initial
exploration of some of the
basic mechanisms.

In \S 2, we introduce the physical model,
derive an equation for the evolution of the mass to magnetic
flux ratio in a weakly ionized medium,
estimate relevant timescales, and establish an initial condition.
In \S 3, we estimate the
turbulent diffusion rate based on mixing length theory,
quantify this estimate with an exact calculation of transport by a wave,
 and argue that
enhanced diffusion requires the introduction of small scales as well as bulk
advection.
In \S 4 we develop a model
based on exponential stretching and shrinking. This model leads to a flux
redistribution rate which is comparable to the eddy turnover rate, as
expected in turbulent diffusion problems. However, the model is
two dimensional, and the fluid motions are prescribed without specifically
allowing for magnetic forces. In \S 5, we consider magnetic feedback.
In \S 6, we apply the model to
the interstellar medium, and discuss the astrophysical constraints imposed by
the dynamics.
Section 7 is a summary and conclusion. Sections 3 and
4 are relevant to general mixing problems, such as turbulent diffusion of a
passive scalar, and the reader who is mainly interested in the astrophysical
implications could go directly to \S 6.

\section{Basic Equations and Model}

\subsection{Equations for Ambipolar Drift}

We consider a
weakly ionized medium with magnetic field $\mathbf{B}$,
mass density $\rho$, and ion mass
density $\rho_i\ll\rho$. We are interested in timescales much longer than the
ion-neutral collision time $\tau_{in}$, in which case
the ion-neutral drift $\mathbf{v_D}\equiv\mathbf{v_i-v_n}$ is well
approximated by
\begin{equation}\label{vd}
\mathbf{v_D}=\frac{\left(\mathbf{\nabla\times B}\right)\mathbf{\times B}}{4\pi
\rho_i}\tau_{in},
\end{equation}
(Shu 1983).

The magnetic field evolves according to the magnetic induction
equation for a perfectly conducting medium
\begin{equation}\label{in}
\frac{\partial\mathbf{B}}{\partial t}=\mathbf{\nabla\times\left ( v_i\times
B \right )},
\end{equation}
Replacing
$\mathbf{v_i}$ with $\mathbf{v_D}+\mathbf{v_n}$, and approximating $\mathbf{v_n
}$ by the center of mass velocity $\mathbf{v}$, we rewrite the induction
equation as
\begin{equation}\label{in2}
\frac{\partial\mathbf{B}}{\partial t}=\mathbf{\nabla\times\left( v\times B
\right)}
+\mathbf{\nabla\times\left( v_D\times B\right)},
\end{equation}
where $\mathbf{v_D}$ is given by eqn. (\ref{vd}). The first term on the RHS of
eqn. (\ref{in2}) can be expanded using the identity $\mathbf{\nabla\times( v\times B)}=\mathbf{B\cdot\nabla v-v\cdot\nabla B-B\nabla\cdot v}$. Then, using the
continuity equation
\begin{equation}\label{continuity}
\frac{\partial\rho}{\partial t}=-\mathbf{v\cdot\nabla}\rho-\rho\mathbf{\nabla
\cdot v}
\end{equation}
we derive an evolution equation for the magnetic field to density ratio
$\mathbf{B}/\rho$
\begin{equation}\label{Brho}
\frac{\partial}{\partial t}\frac{\mathbf{B}}{\rho} + \mathbf{v\cdot\nabla}
\frac{\mathbf{B}}{\rho}=\frac{\mathbf{B}}{\rho}\cdot{\mathbf{\nabla v}} +
\frac{1}{\rho}\mathbf{\nabla\times\left( v_D\times B\right)}.
\end{equation}
The left hand side of eqn.
(\ref{Brho}) is the comoving, or convective, time derivative of $\mathbf{B}/\rho$. The
first term on the right hand side represents stretching of the fieldlines,
and is a consequence of the frozen flux condition. The second term on the
right hand side represents the evolution of $\mathbf{B}/\rho$
caused by ambipolar
drift.

We now restrict ourselves to two dimensional,
incompressible flows perpendicular to a straight
magnetic field. This geometry captures the main effects under study, and is
consistent with the nature of turbulence in a strong, well ordered magnetic
field (Strauss 1976, Sridhar \& Goldreich 1994,
Goldreich \& Sridhar 1997). We recognize that interstellar turbulence is
frequently observed to
be highly supersonic, and thus cannot be entirely
incompressible. We expect that compressible turbulence to
result in magnetic flux transport just as incompressible turbulence does, but
that magnetic feedback on the turbulence is stronger in the compressible case.

Under the assumptions of incompressibility and two dimensionality, the line
stretching term
vanishes, and eqn. (\ref{Brho}) reduces to
\begin{equation}\label{mixing}
\frac{\partial}{\partial t}\frac{B}{\rho} + \mathbf{v\cdot\nabla}
\frac{B}{\rho}=
\frac{1}{\rho}\mathbf{\nabla\cdot}\frac{B^2}{4\pi\rho}\tau_{ni}\mathbf{\nabla}B,
\end{equation}
where $B$ now represents the amplitude of the magnetic field, $\tau_{ni}
\equiv\tau_{in}\rho_n/\rho_i$ is the neutral-ion collision time, and we have
used eqn. (\ref{vd}).
Equation (\ref{mixing}) is close to an advection - diffusion equation for
the flux to mass ratio $Q\equiv B/\rho$
\begin{equation}\label{lmixing}
\frac{\partial Q}{\partial t} + \mathbf{v\cdot\nabla}Q=\frac{1}{\rho}\mathbf{
\nabla\cdot}\lambda\mathbf{\nabla}B,
\end{equation}
provided that we define the diffusivity
$\lambda$ as $\tau_{ni}v_A^2$, where $B/(4\pi\rho)^{1/2}$ is the Alfven speed.

The diffusion of $Q$ is nonlinear in the sense that
$\lambda
=\lambda(Q,\rho)$.  This
nonlinearity can produce sharp fronts along surfaces where $B$ vanishes or is
tightly folded
(Brandenburg \& Zweibel 1994), similar to fronts
created by nonlinear thermal
conduction
(Zel'dovich \& Raizer 1966). Resistive diffusion in these current sheets
alters the mass to flux ratio as well as changing the magnetic topology
(Brandenburg \& Zweibel 1995, Zweibel \& Brandenburg 1997).
In this paper we assume that the relative variation of $B/\rho$ is
so weak that nonlinear effects play only a minor role
in ambipolar drift, and $B$ remains nonsingular.

Equation (\ref{lmixing}) can be used to derive an equation for the rate of
change of $B$ within a comoving volume $V$ of fluid; that is, a fluid element of
fixed mass. We have
\begin{equation}\label{Qdot}
\frac{d}{dt}\int_Vd^3x\rho Q=\int_Vd^3x\frac{\partial}{\partial t}\rho Q +
\int_Sd^2x\rho Q\mathbf{v\cdot\hat n},
\end{equation}
where $S$ is the surface of $V$. The first term on the RHS of eqn.
(\ref{Qdot}) is the Eulerian change of $B$ and the second term accounts for the
motion of $V$. Using eqns. (\ref{continuity}) and (\ref{lmixing}) to expand
the first term on the RHS and applying Gauss' theorem yields
\begin{equation}\label{Qdotf}
\frac{d}{dt}\int_Vd^3x\rho Q=\int_Sd^2x\lambda\mathbf{\hat n\cdot\nabla}B.
\end{equation}
Since the mass within $V$ is constant, eqn.
(\ref{Qdotf}) shows that the flux to mass ratio within a volume moving with
the center of mass velocity decreases
if the magnetic field decreases outward on the surface of the
volume.

\subsection{Timescales}

In the absence of flow, the characteristic diffusion time for a magnetic
field of representative strength $B$ and scale length $L\sim\vert B/\nabla B
\vert$ is
\begin{equation}\label{tdiff}
t_{d0}\equiv\frac{L^2}{\lambda}=\frac{L^2}{v_A^2\tau_{ni}}=\frac{\tau_A^2}{\tau_{ni}},.
\end{equation}
where $\tau_A\equiv L/v_A$ is the global Alfven time.

Expressing $t_{d0}$ in physical units reveals the magnitude of the timescale
problem. We take $\tau_{ni}$ from Draine, Dalgarno, \& Roberge (1983); when
the ratio of ion to neutral atomic weight $A_i/A_n\gg 1$, $\tau_{ni}=6.7\times
10^8 n_i^{-1}$ s. The Alfven speed $v_A=2.2\times 10^5 B_{\mu}/(n_n A_n)^{1/2}$
cm s$^{-1}$. The diffusivity $\lambda$ is then $3.2\times 10^{19}B_{\mu}^2/(A_n
n_n^2 x_i)$ cm$^2$ s$^{-1}$,
where $x_i\equiv n_i/n_n$ is the ionization fraction. The drift
time is
\begin{equation}\label{td0n}
t_{d0}=3.1\times 10^{20}\frac{N_{20}^2 x_i A_n}{B_{\mu}^2}\rm{s},
\end{equation}
where $N_{20}\equiv n_n L$ is the column density in units of 10$^{20}$ cm$^{-2}$. For example, the systems
reported by Heiles (2001) have $N_{20}$ a few tenths to a few,
with $B_{\mu}$ typically 3. If we take $A_n = 1.4$ for gas of cosmic
composition and $x_i = 10^{-4}$, which is probably a conservative lower limit,
then we find $t_{d0}$ is of order 10$^8$ yr for these systems. Since this is
much more than the
10$^6$ - 10$^7$ yr expected lifetime of an interstellar cloud, the diffusion
rate must be enhanced by  a factor of 10 - 100 in order to explain the flatness
of the $B - n$ relation.

Mixing by eddies in the neutral gas requires that the magnetic field be frozen
to the flow. The degree of freezing is measured by the ambipolar Reynolds
number $R_{AD}$, which is large under frozen in conditions
 (Zweibel \& Brandenburg 1997). For eddies of characteristic
size $l$ and speed $v_t$,
\begin{equation}\label{RAD}
R_{AD}\equiv\frac{lv_t}{\lambda}.
\end{equation}
Thus, the field is frozen to the turbulent flow for
\begin{equation}\label{frozen}
\left(\frac{l}{L}\right)\left(\frac{v_t}{v_A}\right) > \frac{\tau_{ni}}{\tau_A}.\end{equation}
Since $\tau_{ni}/\tau_A$ is expected to be small, while $v_t/v_A$ is order
unity, eqn. (\ref{frozen}) implies a fair degree of dynamic range for the
sizes of eddies that can mix the field.
Equation (\ref{RAD}) can also be written as
\begin{equation}\label{RAD2}
R_{AD}=\frac{v_t^2}{v_A^2}\frac{\tau_d}{\tau_{ni}},
\end{equation}
which shows that if the flow is at or above equipartition with the field
($v_t\ge v_A$), the field is frozen in if the neutral-ion collision time is
less than the eddy turnover time.

\subsection{Initial Condition}

It will be useful in the following analysis to have a definite model for $B$.
We take as an initial condition
\begin{equation}\label{ic}
B(x,y,0)=B_{00} + \frac{1}{2}B_0^{''}x^2,
\end{equation}
where $B_{00}$
and $B_0^{''} < 0$ are constants. We will define $\lambda$ using $B_{00}$ for
$B_0$ in quantitative examples.
We view eqn. (\ref{ic}) as the first
two terms in a Taylor expansion of a magnetic field which peaks at $x=0$, and
will assume $x/L\ll 1$.

We first consider pure diffusion.
Motivated by the initial condition
(\ref{ic}), we seek solutions of eqn. (\ref{lmixing}), with $\mathbf{v}=0$, of
the form
\begin{equation}\label{ansatz}
B(x,y,t)=B_0(t) + \frac{1}{2}B_0^{''}x^2.
\end{equation}
Substituting eqn. (\ref{ansatz}) into eqn. (\ref{lmixing}) and using eqn.
(\ref{ic}), we find
\begin{equation}\label{odiff}
B_0(t)=B_{00}+B_0^{''}\lambda t.
\end{equation}
Equation (\ref{odiff}) predicts that the peak field decreases by a factor of
two on a timescale
\begin{equation}\label{td0}
t_{d0}\equiv -B_{00}/(2\lambda B_0^{''}).
\end{equation}
Equation (\ref{td0})
agrees with eqn. (\ref{tdiff})
if we define the magnetic lengthscale $L$ by
\begin{equation}\label{L}
L=\left(-\frac{B_{00}}{2B_0^{''}}\right)^{1/2}.
\end{equation}
\section{Diffusion in the Presence of Waves}

We begin with a mixing length argument.
Consider a magnetic
flux tube of width $a$ which is carried by
a random flow $u$ a distance $l$ down the gradient of
$B$. The field in the tube diffuses into the ambient medium;
thus the motion causes net transport of $B$. The transport is most
effective when
the diffusion time across the tube is comparable to the advection time
\begin{equation}\label{comp}
\frac{a^2}{\lambda}\sim\frac{l}{u},
\end{equation}
because if
$a^2/\lambda\gg l/u$, the flux tubes return to their original positions
with nearly the same value of the field, while if $a^2/\lambda\ll l/u$ the
field diffuses too quickly to be advected by the flow.

Advection spreads $B$ over a distance $l$ in a time $l/u$. In this time,
$B$ would spread diffusively over a distance $(\lambda l/u)^{1/2}$.
By eqn. (\ref{comp}), this distance is just $a$. Advective mixing accelerates
the transport of $B$ only if $a/l < 1$, meaning that the motion consists of
thin fingers which travel much further than their widths (see Ottino 1989 for
discussion of mixing by fingers, or tendrils).
Such fingers are not
seen in models of Alfvenic turbulence, and it is not clear that they would form
in a weakly stratified gas such as the interstellar medium.
Interpreted more broadly, the argument presented here shows that turbulent
mixing requires more than just advection and dispersal; it also requires the
formation of small scales.

Now, we quantify this result.
Weak turbulence theory, in which turbulence is modelled as a superposition of
randomly phased waves, can be used to compute the rate of turbulent diffusion
(eg. Moffatt 1978, Gruzinov \& Diamond 1994). In this so-called quasilinear
 approach, one partitions
quantities into mean and fluctuating parts and calculates the average effect of the fluctuations on the mean part. We used this method in
a previous study of ambipolar diffusion (Zweibel 1988). In the present problem
it is possible, as well as instructive,
 to solve the induction equation exactly instead of averaging
it. This confirms the argument given in \S 2.

We introduce a periodic flow in the $\hat x$ direction
\begin{equation}\label{wave}
\mathbf{v}=\hat x u\sin\omega t\sin ky.
\end{equation}
Motivated by the initial condition eqn. (\ref{ic}), we try a solution of the
advection-diffusion equation (\ref{lmixing}) of the form
\begin{equation}\label{wavesol}
B(x,y,t)=B_0(t) + \frac{1}{2}B_0^{''} + B_1(t)x\sin ky + B_2(t)\cos 2ky.
\end{equation}
Substituting eqn. (\ref{wavesol}) into eqn. (\ref{lmixing}), using eqn. (\ref{wave}),
and equating like powers of $x$ and Fourier harmonics of $y$ leads to a set of
coupled ODEs for the functions
$B_0$, $B_1$, and $B_2$. The solution for $B_0$, which follows the decay of the
peak magnetic field, is
\begin{eqnarray}\label{B0result}
B_0& = &B_{00} + \lambda B_0^{''}t\nonumber \\
    & &+ \frac{B_0^{''}u^2}{2\left(\omega^2+\Gamma^2\right)}
    \Bigg[
										\frac{\Gamma}{2}\left(t - \frac{\sin 2\omega t}{2\omega}\right)
                                               + \frac{\sin^2\omega t}{2}\nonumber \\
  & &+\frac{\omega^2}{\omega^2+\Gamma^2}
                         \left(1-e^{-\Gamma t}\cos\omega t\right)\nonumber \\
  & &-\frac{\omega\Gamma}{\omega^2+\Gamma^2}e^{-\Gamma t}\sin\omega t
       \Bigg],
\end{eqnarray}
%
where $\Gamma\equiv\lambda k^2$.

The maximum decay rate
occurs when the motion given by eqn. (\ref{wave}) is
coherent over many wave periods. The long time behavior of
$B_0$ is then given by
\begin{equation}\label{longtime}
B_0=B_{00} + \left(\lambda + \lambda_t\right) B_0^{''}t,
\end{equation}
where
\begin{equation}\label{lambda_t}
\lambda_t\equiv\frac{\Gamma u^2}{4\left(\omega^2+\Gamma^2\right)}
\end{equation}
represents diffusion brought about by advective transport. Equation (\ref{lambda_t}) closely resembles the turbulent diffusivity calculated from quasilinear theory (Moffatt 1978).
Maximizing $\lambda_t$ with respect to $k$,
we find that the maximum occurs for $\omega = \Gamma$, as we asserted in the mixing length argument following eqn. (\ref{tdiff}), and
is
\begin{equation}\label{lambda_tmax}
\lambda_{t,max}=\frac{u^2}{8\lambda k^2},
\end{equation}
where we have replaced $\Gamma$ by $\lambda k^2$. If we express $u$ in terms
of the maximum fluid displacement $a\equiv u/\omega$ and take the ratio of
$\lambda_{t,max}$ to $\lambda$, the result is
\begin{equation}\label{ratio1}
\frac{\lambda_{t,max}}{\lambda}=\frac{k^2a^2}{8}.
\end{equation}
Equation (\ref{ratio1}) shows that the diffusion rate is appreciably
enhanced by waves only if $ka\gg 1$, meaning
that the flow consists of long, thin streamers (see also Press \& Rybicki
1981). We reached the same conclusion
from mixing length theory. The missing
ingredient is stretching and shrinking of scales, an intrinsic feature of
turbulent flows which we incorporate in the next section.

\section{Stagnation Point Flow}

Hyperbolic stagnation point flow is a particularly tractable example of a flow
with  exponential shrinking and stretching. At hyperbolic stagnation points,
the fluid flow converges in one
(or two) directions and diverges in the other direction(s), while maintaining
incompressibility.
It is well known that diffusion is accelerated in the vicinity
of stagnation points, due to the shrinking of scales in the convergent
directions (Moffatt 1978, Zweibel 1998), while Zel'dovich et al.
(1984) demonstrated dynamo action
by a random
ensemble of stagnation points.
The role of hyperbolic stagnation
points in
the mixing of scalar fields in turbulent flows has recently been reviewed
(Shraiman \& Siggia 2000) with emphasis on the development of intermittency,
and the high order moments of the distribution of concentrations.

The advection-diffusion problem for
fields of the form
(\ref{ic}) is exactly soluble
for stagnation point flow.
In subsection (\ref{singlesp}) we compute the effect of a single stagnation
point. In subsection (\ref{multisp}) we superimpose the effects of a random
ensemble of stagnation points, and in subsection (4.3) we discuss the
relationship between the
stagnation point model and turbulent flow. Our model is not intended to be
a full theory of turbulence, but merely illustrative.

\subsection{A Single Stagnation Point} \label{singlesp}

We consider two dimensional, incompressible flow near a stagnation point at
$(a_x,a_y)$. For fields of the form given by eqn. (\ref{ic}),  we require
$a_x/L\ll 1$, where $L$ is given by eqn. (\ref{L}).  The flow is
\begin{eqnarray}\label{sp1}
v_x=-\gamma\left(x-a_x\right),\\
v_y=\gamma\left(y-a_y\right),
\end{eqnarray}
where $\gamma$ is a constant. It is straightforward to integrate equations
(\ref{sp1}) to find the position at time $t$ of a fluid parcel which is at
position $(x_0,y_0)$ at $t=0$
\begin{eqnarray}\label{pos1}
x=a_x+\left(x_0-a_x\right)e^{-\gamma t}\\
y=a_y+\left(y_0-a_y\right)e^{\gamma t}.
\end{eqnarray}
The
initial coordinates in terms of the coordinates at time $t$ are
\begin{eqnarray}\label{opos1}
x_0=a_x+(x-a_x)e^{\gamma t}\\
y_0=a_y+(y-a_y)e^{-\gamma t}.
\end{eqnarray}

We now compute the effect of this stagnation point flow on the diffusion of the
magnetic field. With eqn. (\ref{ic}) as the initial condition, we look for a
solution of the form
\begin{equation}\label{Bsp1}
B(x,y,t)=B_0(t) + \frac{1}{2}B_0^{''}x_0^2,
\end{equation}
where $x_0(x,y,t)$ is given by eqn. (\ref{opos1}). Substituting eqn. (\ref{Bsp1}) into the advection-diffusion equation (\ref{lmixing}) yields
\begin{equation}\label{adsp}
\dot B_0=\lambda B_0^{''}e^{2\gamma t}.
\end{equation}
$B$ is of the form
(\ref{Bsp1}) exist for two reasons. First, $x_0$ is a
constant of the motion, so for any function $f(x_0)$
\begin{equation}\label{integral}
\frac{\partial f(x_0)}{\partial t} +\mathbf{v\cdot\nabla}f(x_0)=0,
\end{equation}
where $\mathbf{v}$ is given by eqns. (\ref{sp1}). Second,
eqn. (\ref{opos1})
shows that $x_0$ is a linear function of $x$ and $y$, so $\nabla^2 x_0$ is
only a function of time, and is independent of the stagnation point location
$(a_x,a_y)$.

The solution of eqn. (\ref{adsp}) which fits the initial conditions is
\begin{equation}\label{Bsol1}
B_0(t)=B_{00}+\frac{\lambda}{2\gamma}B_0^{''}\left(e^{2\gamma t} - 1\right).
\end{equation}
The location of the peak field evolves in time to $a_x(1-e^{-
\gamma t})$, but this is irrelevant because the density $\rho$ is shifted by
the same amount.
It is only diffusion which affects the mass to flux ratio.

According to eqn. (\ref{adsp}), the peak field has dropped to half its
value in the time $t_{\gamma}$
\begin{equation}\label{tgamma}
t_{\gamma}=\frac{1}{2\gamma}\ln\left(1+2\gamma t_{d0}\right),
\end{equation}
where $t_{d0}$ is the diffusion time in the static case, defined in eqn.
(\ref{td0}). Equation (\ref{tgamma}) shows that the diffusion time depends
only logarithmically on the diffusivity $\lambda$. This arises because of the
exponential growth of the magnetic field gradient, as seen in eqn.
(\ref{adsp}).

The quantity $2\gamma t_{d0}$ which appears in the logarithm in eqn.
(\ref{tgamma}) is, however, a large number. If we identify $\gamma^{-1}\sim
l/v_t$ with an eddy turnover time $\tau_d$, and use eqns. (\ref{tdiff}) and
(\ref{RAD}), we can write
\begin{equation}\label{lp}
2\gamma t_{d0}\sim 2\frac{\tau_A^2}{\tau_d\tau_{ni}}\sim 2 R_{AD}\frac{L^2}{l^2}.
\end{equation}

At the time $t_{\gamma}$, the $\hat x$ component of the drift velocity $v_{Dx}$
is
half the flow velocity $v_x=\gamma x$. Beyond this time, the magnetic field
is not well coupled to the flow.

\subsection{An Ensemble of Stagnation Points}\label{multisp}

We compute the evolution of the magnetic field carried by Lagrangian fluid
elements under
sequences of stagnation point flows oriented
in random directions, each one of which endures for a time $\tau$ (see
Childress \& Gilbert 1995 for a general discussion of these so-called
renewing flows).

We take the flow during time $(n-1)\tau < t < n\tau$ to be
\begin{eqnarray}\label{spr}
v_{nx}=\gamma\mu_n x + \gamma\left(1-\mu_n^2\right)^{1/2}y,\\
v_{ny}=\gamma\left(1-\mu_n^2\right)^{1/2}x-\gamma\mu_n y,
\end{eqnarray}
where $-1\le\mu_n\le 1$ and $ n\ge 1$, and for simplicity we have taken all
flows to have the same strength $\gamma$.
Equation (\ref{spr}) reduces to eqn. (\ref{sp1}) if $\mu=-1$. These flows are
curl free, and hence not of the most general possible type. However,
in the neighborhoods of stagnation points,
vorticity
leads to changes of scale at an algebraic
rather than exponential rate, complicating the mathematics while having
little
effect on the diffusion rate (Zel'dovich et al. 1984, Zweibel 1998). Thus,
vorticity is of secondary importance to our problem, and we omit it here,
although interstellar turbulence almost certainly possesses vorticity.

Let the $(x,y)$ coordinates of a fluid parcel at time $(n-1)\tau$ be $\mathbf{r_{n-1}}$.
Then at time $n\tau$, the coordinates $\mathbf{r_{n}}$ can be written as
\begin{equation}\label{rn}
\mathbf{r_{n}}=\mathbf{A_n\cdot r_{n-1}},
\end{equation}
where the matrix $\mathbf{A_n}$ is
$$\left(\begin{array}{cc}
\cosh\gamma\tau + \mu_n\sinh\gamma\tau & \left(1-\mu_n^2\right)^{1/2}\sinh\gamma\tau \\
\left(1-\mu_n^2\right)^{1/2}\sinh\gamma\tau & \cosh\gamma\tau - \mu_n\sinh\gamma\tau
\end{array}\right)
$$
Inverting eqn. (\ref{rn}) yields $\mathbf{r_{n-1}}$ in terms of $\mathbf{r_{n}}$
\begin{equation}\label{rn-1}
\mathbf{r_{n-1}}=\mathbf{A_n^{-1}}\mathbf{\cdot r_{n}}.
\end{equation}
Successive backwards iteration of eqn. (\ref{rn-1}) yields the initial position
$\mathbf{r_0}$ in terms of the coordinate $\mathbf{r_{N}}\equiv\mathbf{r}(N\tau
,\mathbf{r_0})$
\begin{equation}\label{r0}
\mathbf{r_0}=\mathbf{A_1}^{-1}\cdot\mathbf{A_2}^{-1}...\mathbf{A_N}^{-1}\cdot\mathbf{
r_{N}}.
\end{equation}
At times $n\tau < t < (n+1)\tau$, the position $\mathbf{r}(t)$ is related to the
postion at time $n\tau$ by an equation similar to eqn. (\ref{rn}), where in the matrix
$\mathbf{A}$ we replace $\gamma\tau$ by $\gamma(t-n\tau)$.

Since the
matrices $\mathbf{A_n}$ are independent of the
spatial coordinates,
$\mathbf{r_0}$ is a linear function of $\mathbf{r_{N}}$, or, more generally,
$\mathbf{r}(t)$.
This means that $\nabla^2 x_0^2$ is a function only
of time. It follows that
the solution $B$ of the advection-
diffusion equation (\ref{lmixing}) can still be written in the form of eqn.
(\ref{Bsp1}), and that the rate of diffusion increases with time at the
same rate as
$\nabla^2 x_0^2$.

In order to estimate the rate of increase of $\nabla^2 x_0^2$, we
generated random sequences of $\mu_n$
and calculated
$\mathbf{r_0}$ from $\mathbf{r_N}$ according to eqn. (\ref{r0}).
We used the result to calculate $\nabla^2 x_0^2$ as a function of $n$, or
equivalently, as a function of time, since $n$ corresponds to the time
$n\tau$.  The only adjustable parameter in these calculations is $\gamma\tau$,
which measures the coherence of the flow, here the renewal interval in units of
the stretching rate. We expect $\gamma\tau$ to be $\mathcal{O}(1)$.

Figure (1) shows the dimensionless diffusion rate as a function of iteration
number in the case
$\gamma\tau = 0.5$.
\begin{figure}[!ht]
\begin{center}
\epsscale{.7}
\plotone{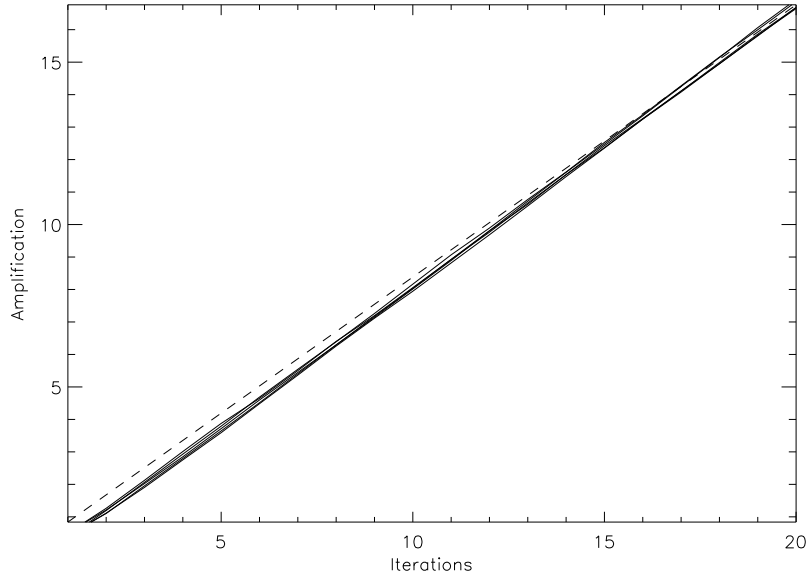}
\end{center}
\caption{The natural log of the
increase in diffusion rate, or amplification factor, with
time, or the number of iterations. Each solid curve is the average of 50
independent realizations of the iteration process. The
dashed line, which has slope 0.838,
is the exponential with the same final value as the average of the curves.
The maximum
amplification factor possible would occur if all stagnation points had inflow
along the $\hat x$ axis, and would have a slope of 1 in these units.}\label{0.5avglog}
\end{figure}
On average, the increase in diffusion rate is well fit by an exponential, and
is about 84\% the rate of increase for coherent stagnation point flow given in
eqn. (\ref{sp1}). After 20 renewals the average diffusion rate is more than
10$^7$ times larger than its original value. This is
much larger than the 2-3 orders of magnitude that we estimated in \S 2.2
as required to explain the
$B-n$ relation.
The mean amplification rate is relatively
insensitive to the coherence parameter $\gamma\tau$, being about 75\% of
maximum if $\gamma\tau = 0.1$ and about 87\% of maximum if $\gamma\tau = 1$.

However, there is substantial
dispersion about the mean. Each solid curve represents 50 independent
sequences of iterations, and Fig. (1)
shows differences between them. The standard
deviation within each set of 50 sequences is typically about 40\%  of the
mean, and the amplification factors for single sequences rarely grow
exponentially. This is illustrated in Figure (2).
\begin{figure}[!ht]
\begin{center}
\epsscale{.7}
\plotone{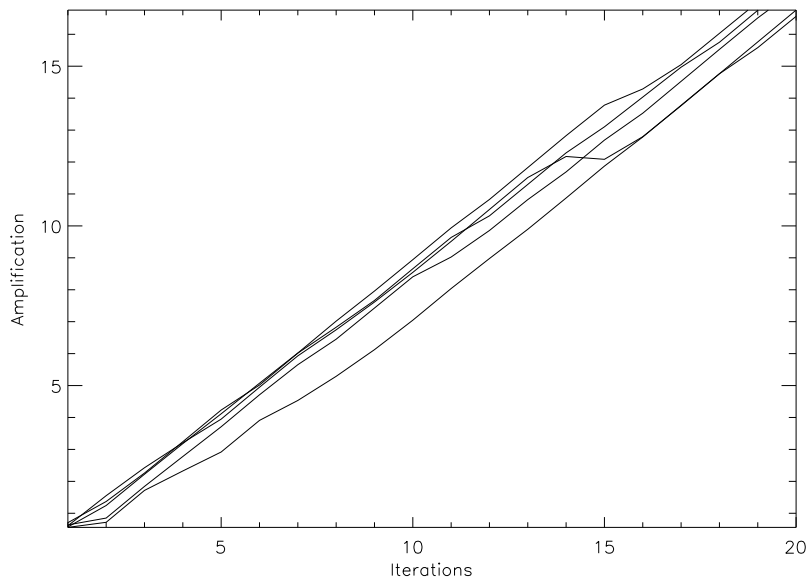}
\end{center}
\caption{The natural log of the
increase in diffusion rate, or amplification factor, with
time, or the number of iterations. Each solid curve is a single
independent realization of the iteration process. The curves, which correspond
to the first 5 members of a larger ensemble,
show the intrinsic variability of the amplification process.}\label{0.5log}
\end{figure}
The large standard deviation suggests that the diffusion rate in this model
is highly
intermittent, which is characteristic of turbulent mixing. Additional evidence
of intermittency is seen in the PDFs of the distribution of amplification
factors, shown in Figure (3) for two different values of $\gamma\tau$. The
maximum possible amplification rate of 20 imposes a cutoff on the high
amplification side of the curve for $\gamma\tau = 1.0$; the PDF for the case
$\gamma\tau = 0.1$ is more symmetrical because the mean amplification rate is
well below the maximum.

\begin{figure}[!ht]
\begin{center}
\epsscale{.7}
\plotone{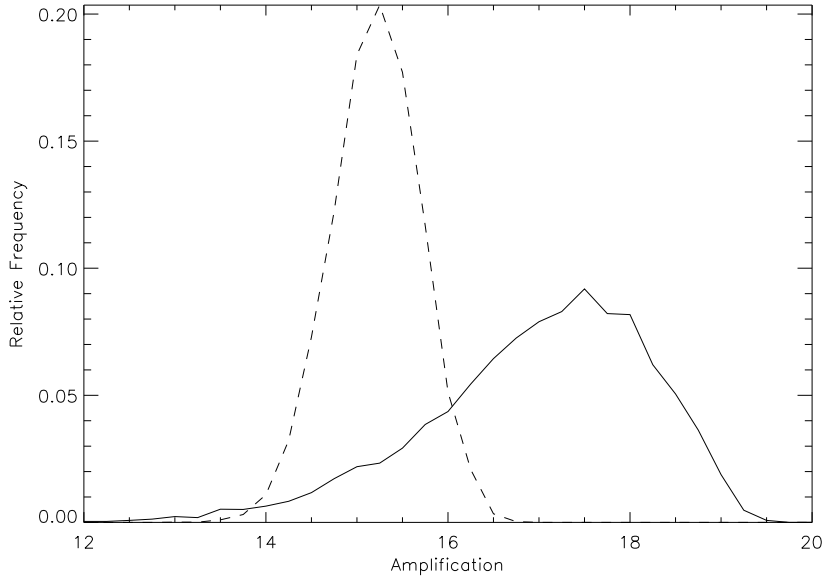}
\end{center}
\caption{Normalized frequency distributions, or PDFs, of the natural logarithms
of the amplification factors for 10 iterations at $\gamma\tau = 1.0$
(solid curve) and 100 iterations at
$\gamma\tau = 0.1$ (dashed curve), so that the curves correspond to the same
total time. Each curve is based on 10$^4$ random sequences of iterations.
}\label{pdfs}
\end{figure}

\subsection{The Stagnation Point Model and Turbulent Flow}

The hyperbolic stagnation point model achieves fast diffusion by
increasing the magnetic field gradient at, on average, nearly exponential
rates. Chaotic flow
achieves fast stretching and shrinking
without hyperbolic stagnation points
because
the trajectories of neighboring fluid particles separate at an exponential rate.
Small differences in the rates of exponential change
lead to highly intermittent
distributions of scalar quantities, which we saw reflected in the stagnation
point model through the wide dispersion of amplification rates (see Figures
 2 \& 3).
The maximum rate of stretching of a fluid element
labelled by its initial position $\mathbf{x_0}$ is given
by the Lyapunov
exponent $\Lambda(\mathbf{x_0})$
\begin{equation}\label{lyap}
\Lambda\left(\mathbf{x_0}\right)\equiv \max_{\mathbf{e_0}}\limsup_{t\rightarrow
\infty}\frac{1}{t}\ln\mid\frac{\partial\mathbf{x}}{\partial\mathbf{x_0}}\cdot
\mathbf{e_0}\mid,
\end{equation}
where the $\mathbf{e_0}$ are the set of all possible unit vectors
(see
Childress \& Gilbert 1995).

In order to make it plausible that the diffusion rate is enhanced by
exponential shrinking and stretching in a flow, we imagine that the
diffusivity $\lambda$ is so small that we can ignore it. In this limit,
the solution of the advection equation for the initial condition (\ref{ic}) is
\begin{equation}\label{cauchy}
B\left(\mathbf{x},t\right)=B\left(\mathbf{x_0},0\right)=B_{00}+\frac{1}{2}
B_{0}^{\prime\prime}x_0^2\left(\mathbf{x}\right).
\end{equation}
If we restore diffusion, it appears in the advection-diffusion
equation (\ref{lmixing}) as the term
\begin{equation}\label{action}
\lambda B_{0}^{\prime\prime}\nabla^2\frac{x_0^2}{2}=\lambda B_{0}^{\prime\prime}\left(x_0\nabla^2 x_0 +\nabla x_0\cdot\nabla x_0\right).
\end{equation}
The quantity $\nabla x_0\cdot\nabla x_0$
on the right hand side of eqn. (\ref{action})
 is the square of the inverse of the stretching rate. This suggests
heuristically that the diffusion rate grows exponentially.

Examples of 2D, chaotic, spatially periodic flow with a single
lengthscale, including maps of Lyapunov exponents and other measures of
chaos, are given by
Galloway \& Proctor (1992), Ponty et al. (1993) and Cattaneo et al. (1995).

\section{Dynamical Feedback with Extension to 3D}

In \S 4,
we prescribed a strictly 2D flow, and neglected
feedback by
Lorentz forces. In fact, by virtue of eqn. (1),
the diffusion rate cannot be enhanced without
increasing the Lorentz force, while 3D effects modify the diffusion process
itself (compare eqns. (3) and (6)).
In this section we quantify the effects of magnetic feedback
and derive criteria for the validity of the 2D model.

We assume the
stagnation point flow given by eqns. (\ref{sp1}) with $\mathbf{a}=0$.
The magnetic field can be written using eqns. (\ref{opos1}), (\ref{Bsp1}),
and (\ref{Bsol1}) as
\begin{equation}\label{b}
B=B_{00}\left[1-\frac{1}{4\gamma t_{d0}}\left(e^{2\gamma t}-1\right) -
\frac{x^2}{4L^2}e^{2\gamma t}\right].
\end{equation}
We evaluate the importance of feedback by following the
force on a fluid element over the mixing time $t_{\gamma}$. This probably
overestimates the effects of magnetic feedback, because the coherence time
$\tau$ of any particular realization of the flow is expected to be less than
$t_{\gamma}$. Thus, the constraints on the turbulence which we derive are
likely to be conservative. In fact, mixing appears to take place in fully
self consistent models of Alfven wave turbulence; Maron \& Goldreich (2001).

In what follows, it is useful to remember that the parameter $2\gamma t_{d0}$,
the ratio of the classical ambipolar diffusion time to the eddy turnover time,
is large; see eqn. (\ref{lp}). We will sometimes use the inverse of this
quantity as an expansion parameter.

\subsection{Magnetic Pressure Forces}

We  estimate the deceleration of an
element of fluid by magnetic pressure in time $t_{\gamma}$. The $\hat x$
component of
magnetic pressure force $F_m$ is
\begin{equation}\label{Fm0}
F_m=-\hat x\frac{\partial}{\partial x}\frac{B^2}{8\pi}.
\end{equation}

The deceleration $\Delta v^P$ of a fluid element over a time $t$ is
\begin{equation}\label{deltavP}
\Delta v^P\left(x_0\right) = \frac{1}{\rho}\int_{0}^{t}ds F_m\left(x(s),s\right),
\end{equation}
where $x(s)=x_0e^{-\gamma s}$ is the position of the fluid element at time $s$
and $x_0$ is its original position.
Substituting eqn. (\ref{b}) into eqn. (\ref{Fm0}) and integrating
eqn. (\ref{deltavP}) to $t=
t_{\gamma}$ yields to leading order in $2\gamma t_{d0}$
\begin{equation}\label{deltavP2}
\Delta v^P\left(x_0\right)=-\frac{B_{00}^2x_0}{16\pi\rho L^2\gamma}\left(2\gamma
t_{d0}\right)^{1/2},
\end{equation}
where we have assumed $x_0/L\ll 1$.

The average velocity $\bar v(x_0)$ of the fluid element over this time is
\begin{equation}\label{vbar}
\bar v\left(x_0\right)\equiv\frac{1}{t_{\gamma}}\int_{0}^{t_{\gamma}}ds
v(x(s),s),
\end{equation}
where $x(s)$ is once again the Lagrangian position of a fluid element. For the
stagnation point flow (\ref{sp1}),
\begin{equation}\label{vbar2}
\bar v =\frac{x_0}{t_{\gamma}},
\end{equation}
to leading order in $2\gamma t_{d0}$. Magnetic feedback on the flow is
unimportant if $\Delta v^P/\bar v < 1$.
 Combining eqns. (\ref{deltavP2}) and
(\ref{vbar2}), we derive
\begin{equation}\label{deltavPvbar}
\frac{\Delta v^P}{\bar v}=\frac{B_{00}^2}{32\pi\rho L^2\gamma^2}\left(2\gamma
t_{d0}\right)^{1/2}\ln\left(2\gamma t_{d0}\right).
\end{equation}
Using eqns. (\ref{RAD}) and (\ref{lp}), eqn. (
\ref{deltavPvbar}) can be rewritten as
\begin{equation}\label{dvPvbar2}
\frac{\Delta v^P}{\bar v}=\frac{\tau_d}{\tau_A}\left(\frac{\tau_d}{32\tau_{ni}}
\right)^{1/2}\ln\left(2\frac{\tau_A^2}{\tau_d\tau_{ni}}\right).
\end{equation}
Equation (\ref{dvPvbar2}) implies an upper limit $\tau_{max}^{P}$ on the eddy
turnover time $\tau_d$ such that $\Delta v^P/\bar v \le 1$;
\begin{equation}\label{taumaxP}
\frac{\tau_{max}^P}{\tau_A}
=\left(18\frac{\tau_{ni}}{\tau_A}\right)^{1/3}\left[\ln\left(
\frac{\tau_A}{\sqrt{2}\tau_{ni}}\right)\right]^{-2/3},
\end{equation}
where, to sufficient accuracy, we have replaced $\tau_d$ by $(32\tau_A^2
\tau_{ni})^{1/3}$ in the logarithmic factor. On this timescale the magnetic
field is still well frozen to the eddies; from eqns. (\ref{RAD}) and
(\ref{taumaxP}),
\begin{equation}\label{RADPmax}
R_{AD}\left(\tau_{max}^P\right)=\frac{v_t^2}{v_A^2}\left(18\frac{\tau_A^2}{\tau_{ni}^2}\right)^{1/3}\left[\ln\left(\frac{\tau_A}{\sqrt{2}\tau_{ni}}\right)\right]^{-2/3}.
\end{equation}

\subsection{3D Effects}

We assume the turbulent motions are in
the $(x,y)$ plane, but depend weakly on $z$; \textit{i.e.} the characteristic
wavenumber $k$ along the field is related to the turbulent lengthscale $l$ by
$kl\ll 1$. This quasi-two dimensionality is expected to be a feature of
Alfvenic turbulence in strong magnetic fields (Strauss 1976, Goldreich \&
Sridhar 1997).

Since the field is fairly well frozen in even on scales $l$, it is
very well frozen on the scale $k^{-1}$, and the transverse field $\mathbf{B_{
\perp}}$ is given to a good approximation by
\begin{equation}\label{Bperp}
\mathbf{B_{\perp}}=B_z(\mathbf{x_0},t)\frac{\partial\mathbf{x_{\perp}}}{\partial z},
\end{equation}
where $\mathbf{x_0}$ is the initial position of the fluid element at position
$\mathbf{x}$ at time $t$.

Let us
introduce a small parameter $\epsilon$ and assume that $\partial_z$ is
$\mathcal{O}(\epsilon)$ relative to the perpendicular derivatives, and that
$B_{\perp}/B_z$ is also $\mathcal{O}(\epsilon)$ (this is the so-called reduced
MHD ordering; Strauss 1976).
It can then be shown
that the changes in the ambipolar drift terms
are of order $\epsilon^2$. Therefore, we
may assume that weak three dimensionality has little effect on
ambipolar drift of the vertical field.

The bent field exerts a tension force which decelerates the fluid
by an amount $\Delta v^T$.
We compute $\Delta v^T$ for a $z$ dependent stagnation point flow model with
\begin{eqnarray}\label{spz}
v_x=-\gamma x\cos kz,\\
v_y=\gamma y\cos kz,
\end{eqnarray}
a generalization of eqns. (\ref{sp1}).
The Lagrangian positions are
\begin{eqnarray}\label{posz}
x=x_0e^{-\gamma t\cos kz},\\
y=y_0e^{\gamma t\cos kz}.
\end{eqnarray}
According to eqns. (\ref{Bperp}) and (\ref{posz}), the $\hat x$ component of
the field is
\begin{equation}\label{Bx}
B_x=B_z\gamma t kx\sin kz.
\end{equation}
The magnetic tension force $F_m$ in the $\hat x$ direction is
\begin{equation}\label{Ft}
F_m=\frac{B_z}{4\pi}\frac{\partial B_x}{\partial z}=\frac{B_z^2}{4\pi}\gamma t
k^2x\cos kz.
\end{equation}
We set $z=0$ and follow a procedure similar to the derivation of eqn.
(\ref{dvPvbar2}), integrating
$F_m$ along the path of a fluid element from $t=0$
to $t=t_{\gamma}$. We approximate $B_z$ by $B_{00}$, which overestimates $F_m$.
The result to leading order in $(2\gamma t_{d0})^{-1}$ is
\begin{equation}\label{dvt}
\Delta v^T=\frac{1}{\rho}
\int_0^{t_{\gamma}}dt F_m = \frac{k^2 v_A^2 x_0}{\gamma}.
\end{equation}
Using eqn. (\ref{vbar2}),
the relative deceleration is $\Delta v^T/\bar v$
\begin{equation}\label{dvvbart}
\frac{\Delta v^T}{\bar v} = \frac{k^2 v_A^2}{2\gamma^2}\ln\left(2\gamma t_{d0}
\right).
\end{equation}
Equation (\ref{dvvbart}) shows that magnetic tension has little effect on the
fluid as long as the Alfven frequency along the fieldline is less than the
eddy turnover rate by the factor $[\ln\left(2\gamma t_{d0}\right)]^{1/2}$.

Equation (\ref{dvvbart})
 can be used to set an upper limit $\tau_{max}^T$ on the eddy
turnover time such that the magnetic field reaches the mixing scale without
decelerating the fluid. Proceeding as in the derivation of eqn. (\ref{taumaxP})
we find
\begin{equation}\label{taumaxT}
\frac{\tau_{max}^T}{\tau_A}=\frac{\sqrt{2}}{kL}\left[\ln\left(\sqrt{2}kL\frac{
\tau_A}{\tau_{ni}}\right)\right]^{-1/2}.
\end{equation}
Equation (\ref{taumaxT}) shows that tension forces are less important in long,
thin structures, in which $kL$ can be much less than one, than they are in
flattened structures such as disks. The field is well frozen to the eddies
as long as $\tau_A/\tau_{ni}\gg 1$.

\section{Application to the Galactic Magnetic Field}

Little is know about interstellar turbulence beyond its gross energetics:
the turbulent kinetic energy is at or above equipartition with the magnetic
energy. Energy injection by a
variety of mechanisms, combined with nonlinear processes, should lead to a
turbulent spectrum over a wide range in scales.

In principle, any weakly ionized interstellar structure which
survives for several eddy turnover times is a
candidate for turbulent ambipolar diffusion.
We showed in \S 5 that the
efficiency of turbulent ambipolar drift can be limited
by the back reaction of magnetic forces. In the strictly 2D case,
the increase in ambipolar diffusion rate is associated with the local buildup
of magnetic pressure forces, and in the 3D case, by
magnetic tension forces as well.
We expressed these constraints in terms of
lower bounds on the strain rates, or inverse eddy turnover times, such that
concentration of the field occurs before deceleration of the flow. These
constraints appear in eqns. (\ref{taumaxP}) and (\ref{taumaxT}). Here,
we express them numerically.

The critical parameters are the ratio $\tau_{ni}/\tau_A$ and the geometrical
factor $kL$. Referring back to \S 2.2 for numerical values, we have
\begin{equation}\label{tratio}
\frac{\tau_{ni}}{\tau_A}=4.8\times 10^{-5}\frac{B_{\mu}}{L_{pc}n_i\left(n_nA_n
\right)^{1/2}},
\end{equation}
where $L$ is expressed in parsecs. For example, if $n_n$ = 50 cm$^{-3}$, $n_i$
= 5 $\times$ 10$^{-3}$ cm$^{-3}$, $A_n$ = 1.4, $L_{pc}$ = 1, $B_{\mu}$ = 3,
$\tau_{ni}/{\tau_A}$ = 3.4 $\times$ 10$^{-3}$, and eqn. (\ref{taumaxP})
requires $\tau_d/\tau_A <$ 0.13. If $v_t\sim v_A$, turbulence on the scale of
a tenth of a parsec or less can mix the magnetic field down to the ambipolar
diffusion scale. On the other hand, eqn. (\ref{taumaxT}) requires
$\tau_d/\tau_A <$ 0.09($2\pi/kL$), which is a more severe constraint,
especially in a highly flattened structure.

There is at least one type of H I structure in which flux freezing appears to
be obeyed.
Magnetic fields in H I shells are observed to be quite strong, with magnitudes
consistent with shock compression (Heiles 1989). The same observations
suggest that turbulence with Alfvenic or slightly subAlfvenic
velocities is present. If these shells were not expanding, they would appear to
fulfill
the conditions for fast ambipolar drift, and their strong fields would be
counterexamples to the theory. However, expansion of the shells at speeds of
order 10 - 20 km/s adds new magnetic flux
faster than it can diffuse upstream, while on the downstream side the
ionization is too high for efficient ambipolar drift. Thus, the field in the shells remains large. Recent observations by
Heiles (2001b) and Heiles
\& Troland (2001) of cold, moderately dense H I regions which have weak magnetic fields, Alfvenic random velocities, and no observed association with shells appear to be better candidates for fast ambipolar drift.

\section{Summary and Conclusions}

Observations show that while interstellar magnetic fieldstrength and
gas density are
to some
extent correlated, the fieldstrength is often lower than expected. This
suggests that processes beyond ideal MHD may play a role.

The flux to mass ratio is altered by ambipolar drift, but
estimates of the ambipolar diffusivity $v_A^2\tau_{ni}$ predict that
ambipolar drift is important only in very dense, strongly magnetized gas with
substantial gradients on small scales.
However, it is well established that turbulence
can enhance the transport rates of quantities such as entropy and angular
momentum. This motivated us to consider the effect
of turbulence on the rate of ambipolar drift. Enhancement by roughly two orders
of magnitude would explain the observations.

As a first attempt on the problem, we considered the geometrically restricted
case of a straight magnetic field, with a transverse gradient, mixed by
2D, perpendicular turbulence. In this situation,
ambipolar drift is described by a nonlinear
diffusion term [eqn. (\ref{mixing})]. For simplicity, however,
we approximated the diffusivity as linear.

We showed by a mixing length argument, and then an explicit
calculation (\S 3) that advection of the field by a periodic flow reduces its
peak value. However, unless the motions are long and thin, like streamers,
the rate of relaxation is no faster than relaxation by ambipolar drift alone.
The missing ingredient is stretching and shrinking of
scales, which in chaotic flows happens at an exponential rate. In \S 4 we
modelled these exponential changes of scale by representing the flow as
a sequence
of randomly
oriented hyperbolic stagnation points. With this model, and a parabolic
initial condition for the magnetic profile,
the advection-diffusion equation (\ref{lmixing}) can be solved exactly. The
model predicts an exponential
increase of the mean diffusion rate with time (Figure 1), although with
considerable variance from point to point (Figures 2 \& 3).
The stagnation point
model predicts that the field diffuses on a timescale comparable to the eddy
turnover time, with only logarithmic dependence
on the ambipolar diffusivity itself
and on the original gradient lengthscale.

In \S 5, we estimated the back reaction of magnetic pressure and
tension forces on the
stagnation point flow, including weak three dimensionality. The relative
deceleration of the fluid over one mixing time is given for pressure forces by
eqn. (\ref{deltavPvbar}) and for tension forces by eqn. (\ref{dvvbart}).
Comparing the deceleration time of
a fluid element to the mixing time, we derived upper limits on the eddy turnover
time such that deceleration is order unity or less within a mixing time
[eqns. (\ref{taumaxP}) and (\ref{taumaxT})].
As we showed in \S 6, these
criteria can be satisfied in the interstellar medium without extreme
assumptions about the size and velocity of the turbulent eddies,
although they cannot be wholly ignored. These
estimates of feedback are
conservative in the sense that
a fully self consistent model of MHD turbulence can still produce fast mixing,
as shown by Maron \& Goldreich (2001) for spreading of a passive scalar. This
conclusion may be dependent on geometry, however,
 as concluded by Kim (1997) based on
a study of turbulent decay
of a \textit{coplanar} magnetic field caused by 2D incompressible motions in
weakly ionized fluid.

The outcome of these calculations is that turbulence is likely to have a major
effect on the magnetic flux to mass ratio in the weakly ionized portions of
the interstellar medium, making the magnetic
field more uniform. The model presented here applies to regions with
simple magnetic topology,
filamentary structure, and no global cross-field flows.
It follows that the strength of the field is not
necessarily a good indicator of the dynamical processes which determine the gas
density.

Although the results presented here are consistent with the conventional
wisdom that turbulent mixing takes place in an eddy turnover time, it is
important to recognize that they are not obtained from a full model of
turbulence. We
plan to extend this work to  more realistic models
which include compressibility, vorticity, and a self consistent
treatment of magnetic forces as well as more general magnetic geometry. If the
stagnation point model holds up in comparison with more complete models, it
could be useful in other mixing problems.

Magnetic reconnection can also change the magnetic flux to mass ratio.
Lazarian \& Vishniac (1999) have argued that if the spectrum of interstellar
turbulence extends to the resistive scale then reconnection takes place at
the Alfven speed. We have concentrated here on ambipolar drift because it
does not require turbulent structure on such small scales; if flux is quickly
redistributed in the fully ionized portions of the ISM then an alternative
process is certainly required.


\acknowledgments

This
work was initiated during the program on Astrophysical
Turbulence at the ITP in Santa Barbara in 2000. I am happy to acknowledge
useful discussions with Nic Brummell, Dick Crutcher, George Field, Carl
Heiles, Fabian Heitsch, David Hughes, Eun-jin Kim, Steve Tobias, Tom
Troland, and especially the referee, Pat Diamond. Material support
was provided by NSF Grants AST 9800616 and AST 0098701 to the University
of Colorado and PHY 9407194 to UC Santa Barbara.

\end{document}